# Velocity-Verlet-like algorithm for simulations of stochastic dynamics

Tobias Gleim

Molecular simulations of many particles which move rather according to a brownian than a newtonian type of dynamics, nevertheless, can be performed by means of a "velocity-Verlet-like" algorithm. The derivation of this algorithm requires the "Ito formula" of stochastic calculus which usually is not part of a scientist's education. Therefore, it is going to be shown, how this formula can be motivated and applied in order to find that algorithm. Furthermore, it is demonstrated, why it is sufficient to use uniformly distributed random numbers within that algorithm, thus avoiding gaussian distributed ones, although gaussian distributed white noise forces are assumed to model the brownian-like motion of the particles. Finally, a solution to the problem is presented that the linear total momentum is not conserved due to the presence of stochastic forces in the equations of motion of the particles.

Newtonian equations of motion of a many-particle system can be solved numerically by means of the famous velocity-Verlet algorithm (see e.g. [1, 2]) being correct including the quadratic order in a time step $h$. But it is less well known that this algorithm can be extended in a minimal way to solve equations of motion of particles obeying something in between a newtonian and a brownian type of dynamics, what we are going to call a "stochastic" type of dynamics in the subsequent text. A benefit of that extended algorithm may be, for instance, that one can perform computer simulations for a canonical ensemble of particles and describe time-dependent thermodynamic properties (e.g. like the mean-squared displacement) of that system: Monte-Carlo methods, e.g., are suitable for the former task, but might not be the best choice for the latter one, whereas Molecular-Dynamics methods like the "pure" velocity-Verlet algorithm are appropriate for the description of time-dependent thermodynamic properties, but only in a microcanonical and not in a canonical ensemble. Having the possibility to describe a mixture between a newtonian and a brownian type of dynamics in computer simulations can be an attractive option when one aims at a comparison with a realistic system, where the kind of dynamics the particles are obeying to might not be obvious. But of course, there are excellent alternatives to the velocity-Verlet-like algorithm presented in the following, e.g. like Nosé-Hoover chains or the related dissipative particle dynamics (see e.g. [2]).

A mathematical treatment for deriving among other things the so-called Heun algorithm, which is taken as basis for that velocity-Verlet-like algorithm, can be found in ref. [3,4]. The algorithm can be read off from the calculations presented e.g. in [5], where the Heun algorithm is used in a much broader context than it is going to be done in this text. Hence, the intention of the subsequent explanations is just to offer an introductory text to facilitate the understanding of that velocity-Verlet-like algorithm, because at first sight, some of its properties seem to be hard to comprehend: for instance, the fact that it is sufficient to apply uniformly distributed random numbers, thus avoiding gaussian distributed ones, although gaussian distributed white noise forces are used to model the brownian-like motion of the particles. That motion can be described by equations for the positions $\vec{r}_j$ and the velocities $\vec{v}_j$ of a particle $j$ with mass $m_j$ in the following way:

$$\dot{\vec{r}}_j = \vec{v}_j,$$ (1a)

$$m_j \dot{\vec{v}}_j = \vec{F}_j - m_j \gamma \vec{v}_j + \sigma_j \vec{\eta}_j(t),$$ (1b)

where the damping constant $m_j \gamma$ is connected with the gaussian distributed white noise force $\sigma_j \vec{\eta}_j(t)$ with zero mean,

$$\langle \vec{\eta}_j(t) \rangle = 0, \tag{2}$$

via the fluctuation dissipation theorem

$$\langle \vec{\eta}_j(t) \cdot \vec{\eta}_l(t) \rangle = 3\delta(t - t')\delta_{jl} \tag{3}$$

with

$$\sigma_j^2 = 2 m_j \gamma k_B T \tag{4}$$

containing the temperature $T$ of a system which, e.g., consists of $N$ particles swimming in a solvent. The external forces $\vec{F}_j$ on a particle $j$ obey Newton's third law, i.e.

$$\sum_{j=1}^N \vec{F}_j = 0. \tag{5}$$

The Lagevin-type of equations (1-4) can be regarded in two different limits. In the limit of strong damping, i.e. $1/\gamma$ being very small, the acceleration term in (1b) can be neglected:

$$\dot{\vec{r}}_j = \frac{1}{m_j \gamma} \vec{F}_j + \frac{1}{m_j \gamma} \sigma_j \vec{\eta}_j(t). \tag{6}$$

These equations are of a brownian type of dynamics, whereas in the limit of a vanishing damping constant, the newtonian equations of motion are reproduced:

$$\dot{\vec{r}}_j = \vec{v}_j, \tag{7 a}$$

$$m_j \dot{\vec{v}}_j = \vec{F}_j, \tag{7 b}$$

Newton's third law (5) leads to the fact that the total linear momentum,

$$\vec{P} = \sum_{j=1}^N m_j \vec{v}_j \tag{8}$$

is conserved for a system of newtonian particles described by (7), but when described by (1) (or (6)), this conservation law is not valid any more. However, the conservation of total linear momentum can be imposed on (1) by the requirement that

$$\sum_{j=1}^N m_j \vec{v}_j(t_0) = 0 \tag{10}$$

at a starting time $t_0$ and by the following restriction on the random forces for all times $t$:

$$\sum_{j=1}^{N} \sigma_j \vec{\eta}_j(t) = 0 \qquad (11)$$

But (11) must be regarded as a condition for the random forces in addition to the fluctuation dissipation theorem, therefore having an influence on the choice of random numbers within a numerical algorithm for solving (1).

The equations of motion (1) and (6) can be reformulated in the subsequent way:

$$\dot{x}^i = f^i(\vec{x}) + \sum_j \sigma^{ij} \eta_j(t) \qquad (12)$$

with

$$\vec{x} = (\vec{r}_1, \ldots, \vec{r}_N, m_1 \vec{v}_1, \ldots, m_N \vec{v}_N) = (x^i)_{i=1,\ldots,6N}, \qquad (13\ \text{a})$$

$$\vec{f}(\vec{x}) = (\vec{v}_1, \ldots, \vec{v}_N, -m_1 \gamma \vec{v}_1 + \vec{F}_1, \ldots, -m_N \gamma \vec{v}_N + \vec{F}_N,), \qquad (13\ \text{b})$$

$$\sigma^{ij} = d_i \delta_{ij}, \quad d_i = \begin{cases} 0, & i=1,\ldots,3N \\ \sigma_i, & i>3N \end{cases}, \quad (i,j=1,\ldots,6N) \qquad (13\ \text{c})$$

for (1) and

$$\vec{x} = (\vec{r}_1, \ldots, \vec{r}_N), \qquad (14\ \text{a})$$

$$\vec{f}(\vec{x}) = \left( \frac{1}{m_1 \gamma} \vec{F}_1, \ldots, \frac{1}{m_N \gamma} \vec{F}_N \right), \qquad (14\ \text{b})$$

$$\sigma^{ij} = \sqrt{\frac{2 k_B T}{m_i \gamma}} \delta_{ij} \qquad (14\ \text{c})$$

for (6). The $\eta_j$ are again gaussian distributed with

$$\langle \eta_j(t) \rangle = 0, \qquad (15\ \text{a})$$

$$\langle \eta_i(t) \eta_j(t') \rangle = \delta_{ij} \delta(t-t'). \qquad (15\ \text{b})$$

In the next chapter, we are going to present numerical algorithms for solving equation (12) by taking into account (15).

**Numerical algorithms**

A numerical solution for the system of stochastic equations (12) can be found by integration over times between 0 and $h$:

$$x^i(h) = x^i(0) + \int_0^h ds\, f^i(\vec{x}(s)) + \sigma^{ij} \int_0^h ds\, \eta_j(s)$$
. (16)

As was done here with the index *j*, we will always sum over same indices. Furthermore, we introduce the integral

$$W_j(h) = \int_0^h ds\, \eta_j(s)$$
. (17)

$f^i$ in (16) can be expanded for small *h*, if (16) is taken again as the argument $\vec{x}$ of $f^i(\vec{x}(s))$ therein. But this expansion cannot be an ordinary Taylor series,

$$f^i(\vec{x}(s)) \approx f^i(\vec{x}(0)) + s\left(\tfrac{d}{ds} f^i\right)_{s=0}(\vec{x}(s)) + \tfrac{1}{2} s^2 \left(\tfrac{d^2}{ds^2} f^i\right)_{s=0}(\vec{x}(s)) + \ldots ,$$
 (18)

because by application of the compound rule of differential calculus in (18), in the third term on the right hand side of (18), derivatives of $\eta_i(t)$ would appear which are not defined (see e.g. [6]). Instead, the so-called Ito-formula must be used. If we had a scalar function *f* of a scalar variable $x(t)$, *f* could be approximated by:

$$f(x(h)) \approx f(x(0) + \Delta x) \approx f(x(0)) + f'(x(0))\Delta x + \tfrac{1}{2} f''(x(0))(\Delta x)^2$$
 (19)

with quantities corresponding to (16) and (17) and assuming $\Delta x$ to be small because the time step *h* is small,

$$\Delta x = x(h) - x(0) = \int_0^h ds\, f(x(s)) + \sigma W(h)$$
, (20 a)

$$W(h) = \int_0^h ds\, \eta(s)$$
 (20 b)

and $x(t)$ obeying a stochastic differential equation similar to (12) and (15):

$$\dot{x} = f(x) + \sigma \eta(t),$$ (21 a)
$$\langle \eta(t) \rangle = 0,$$ (21 b)
$$\langle \eta(t)\eta(t') \rangle = \delta(t - t').$$ (21 c)

Now we have to substitute (20 a) into (19). But we only want to take into account terms of orders including *h*. If the stochastic term in (20 a) or (21 a) were not present, we would not need to consider terms like $(\Delta x)^2$, because the integral term in (20 a) is of order h and therefore $(\Delta x)^2$ would be of order $h^2$ already. But due to the presence of e.g. $W(h)$ in (20 a), we must be more careful:

$$\left\langle W^2(h) \right\rangle = \int_0^h ds \int_0^h ds' \left\langle \eta(s)\eta(s') \right\rangle = \int_0^h ds \int_0^h ds' \, \delta(s-s') = h \tag{22}$$

where we have used (20 b) and (21 c). (22) demonstrates that $W^2(h)$ is just of order $h$ and therefore must be accounted for in (19):

$$f(x(h)) \approx f(x(0)) + f'(x(0)) \left( \int_0^h ds \, f(x(s)) + \sigma W(h) \right) + \tfrac{1}{2} f''(x(0)) \sigma^2 W^2(h) \tag{23}$$

For a non-scalar function $f^i$ of a variable $\vec{x}$, by analogy with (23), we get:

$$f^i(\vec{x}(s)) \approx f^i(\vec{x}(0)) + \left(\partial_m f^i\right)(\vec{x}(0)) \cdot \left[ \int_0^s du \, f^m(\vec{x}(u)) + \sigma^{mn} W_n(s) \right]$$
$$+ \tfrac{1}{2} \left(\partial_p \partial_m f^i\right)(\vec{x}(0)) \sigma^{pq} \sigma^{mn} W_q(s) W_n(s) \tag{24}$$

With the help of a further iteration of (24), we approximate the following term in (24) by

$$\int_0^s du \, f^m(\vec{x}(u)) \approx \int_0^s du \, f^m(\vec{x}(0)) \approx s \, f^m(\vec{x}(0)) \tag{25}$$

Substituting (24) (with (25)) into (16), $x^i(h)$ reads:

$$x^i(h) = x^i(0) + h f^i + \tfrac{1}{2} h^2 f^k \partial_k f^i + \sigma^{ij} W_j(h) + F_l(h) \sigma^{kl} \partial_k f^i + \tfrac{1}{2} G_{qn}(h) \partial_k \partial_m f^i \sigma^{kq} \sigma^{mn} \tag{26}$$

with $f^i \equiv f^i(\vec{x}(0))$,

$$W_i(h) = \int_0^h ds \, \eta_i(s) \tag{27 a}$$

$$F_i(h) = \int_0^h ds \, W_i(s) \tag{27 b}$$

$$G_{ij}(h) = \int_0^h ds \, W_i(s) W_j(s) \tag{27 c}$$

Since $W_i^2(h)$ is of order $h$, $W_i(h)$ is only of order $h^{1/2}$ and therefore $F_i(h)$ and $G_{ij}(h)$ are of order $h^{3/2}$ and $h^2$, respectively. Hence (26) is of order $h^2$. Whereas the functionals $W_i(h)$ and $F_i(h)$ are just sums of gaussian random numbers $\eta(t)$ and therefore are gaussian distributed themselves, this is no longer the case for $G_{ij}(h)$. To get rid of $G_{ij}(h)$ and the derivatives of $f^i$, it is necessary to search for an approximation $\bar{x}^i(h)$ of $x^i(h)$ avoiding these terms. This approximation is acceptable, if it reproduces an observable $\langle M(\bar{x}(h))\rangle$ to the same order in $h$ as (26) does. If the time step $h$ is small,

$$\Delta x^i = x^i(h) - x^i(0) \tag{28}$$

must be small, too (cf. (20 a)). Therefore we can expand $\langle M(\bar{x}(h))\rangle$ to an order including $h^2$ in $x^i(0)$:

$$\langle M(\bar{x}(h))\rangle \approx M(\bar{x}(0)) + (\partial_i M)(\bar{x}(0))\langle \Delta x^i\rangle + \frac{1}{2!}(\partial_i \partial_j M)(\bar{x}(0))\langle \Delta x^i \Delta x^j\rangle$$
$$+ \frac{1}{3!}(\partial_i \partial_j \partial_k M)(\bar{x}(0))\langle \Delta x^i \Delta x^j \Delta x^k\rangle +$$
$$+ \frac{1}{4!}(\partial_i \partial_j \partial_k \partial_l M)(\bar{x}(0))\langle \Delta x^i \Delta x^j \Delta x^k \Delta x^l\rangle \approx M(\bar{x}(0)) + M_1 h + M_2 h^2 \tag{29}$$

To the order $h^2$, inclusively, we get by using (26) and (28):

$$\langle \Delta x^i\rangle \approx f^i h + f^m \partial_m f^i \tfrac{1}{2} h^2 + \partial_m f^i \sigma^{mn}\langle F_n(h)\rangle + \sigma^{ij}\langle W_j(h)\rangle$$
$$+ \tfrac{1}{2}\partial_p \partial_m f^i \sigma^{pq} \sigma^{mn}\langle G_{qn}(h)\rangle \tag{30 a}$$

$$\langle \Delta x^i \Delta x^j\rangle \approx f^i f^j h^2 + \sigma^{ik}\sigma^{jn}\langle W_k(h)W_n(h)\rangle + (f^i \sigma^{jk} + f^j \sigma^{ik})h\langle W_k(h)\rangle$$
$$+ \partial_m f^i \sigma^{mn}\sigma^{jk}\langle F_n(h)W_k(h)\rangle + \partial_m f^j \sigma^{mn}\sigma^{ik}\langle F_n(h)W_k(h)\rangle \tag{30 b}$$

$$\langle \Delta x^i \Delta x^j \Delta x^k\rangle \approx \left[f^i \sigma^{jl}\sigma^{km}h\langle W_l(h)W_m(h)\rangle + \text{permutations of }(i,j,k)\right]$$
$$+ \sigma^{il}\sigma^{jm}\sigma^{kn}\langle W_l(h)W_m(h)W_n(h)\rangle \tag{30 c}$$

$$\langle \Delta x^i \Delta x^j \Delta x^k \Delta x^l\rangle \approx \sigma^{ip}\sigma^{jq}\sigma^{kr}\sigma^{lm}\langle W_p(h)W_q(h)W_r(h)W_m(h)\rangle \tag{30 d}$$

Obviously, we have to calculate seven averages of functionals by considering (15) and (27 a)

$$\langle W_j(h)\rangle = \int_0^h ds \langle \eta_j(s)\rangle = 0 \tag{31 a}$$

because of (15 a). With this result, we can calculate

$$\langle F_j(h)\rangle = \int_0^h ds \langle W_j(h)\rangle = 0 \tag{31 b}$$

by means of (27 b). (15 b) can be used together with (27 a), in order to obtain

$$\langle W_i(h)W_j(h)\rangle = \int_0^h ds \int_0^h ds' \langle \eta_i(s)\eta_j(s')\rangle = \int_0^h ds \int_0^h ds' \delta(s-s')\delta_{ij} = h\delta_{ij}$$
. (31 c)

The white noise terms $\eta_j(t)$ are assumed to be uncorrelated from each other at equal times, only if the indices are different, what can be extracted from (15 b), i.e.:

$$\langle \eta_i(t)\eta_j(t)\rangle = 0, \text{ if } i \neq j.$$

The same is true for a product of more than two $\eta_j(t)$, e.g.:

$$\langle \eta_i(t)\eta_j(t')\eta_k(t'')\rangle = \langle \eta_i(t)\eta_j(t')\rangle \langle \eta_k(t'')\rangle = 0 \qquad (31\ d')$$

if $i \neq k \neq j$ and because of (15 a). Integration of (31 d) over $t$, $t'$ and $t''$ from 0 to $h$ gives

$$\langle W_i(h)W_j(h)W_k(h)\rangle = 0 \qquad (31\ d)$$

due to (27 a). The analogue of (31 d') for a product of four $\eta_j(t)$ reads

$$\langle \eta_i(t_1)\eta_j(t_2)\eta_k(t_3)\eta_l(t_4)\rangle = \langle \eta_i(t_1)\eta_j(t_2)\rangle\langle \eta_k(t_3)\eta_l(t_4)\rangle +$$
$$\langle \eta_i(t_1)\eta_k(t_3)\rangle\langle \eta_j(t_2)\eta_l(t_4)\rangle +$$
$$\langle \eta_i(t_1)\eta_l(t_4)\rangle\langle \eta_j(t_2)\eta_k(t_3)\rangle \qquad , (31\ e')$$

where all possible combinations of two pairs of indices have been considered. Integrating (31 e') over $t_1,...,t_4$ from 0 to h, applying (27 a) and using (31c), yields:

$$\langle W_i(h)W_j(h)W_k(h)W_l(h)\rangle = h^2(\delta_{ij}\delta_{kl} + \delta_{ik}\delta_{jl} + \delta_{il}\delta_{jk}). \qquad (31\ e)$$

Starting from (31e), it is easy to conclude that

$$\langle W_i(h)W_j(h)W_k(h)W_l(h)W_m(h)\rangle = 0, \qquad (31\ f)$$

because (31 f) can be represented by a sum over a product of terms like (31 e) with a single $W_i(h)$. These terms are always zero, by a close analogy with (31 d'). By means of (31c), we get:

$$\langle G_{ij}(h)\rangle = \int_0^h ds \langle W_i(s)W_j(s)\rangle = \int_0^h ds\, h\, \delta_{ij} = \tfrac{1}{2}h^2\, \delta_{ij}. \qquad (31\ g)$$

The calculation of the subsequent average of a functional is a bit more difficult:

$$\langle F_j(h)W_j(h)\rangle = \left\langle \int_0^h ds\, W_i(s)W_j(h)\right\rangle = \int_0^h ds\, \langle W_i(s)W_j(h)\rangle =$$

$$\int_0^h ds \int_0^s dt_1 \int_0^h dt_2 \langle \eta_i(t_1)\eta_i(t_2)\rangle = \int_0^h ds \int_0^s dt_1 \int_0^h dt_2\, \delta(t_1 - t_2)\delta_{ij} =$$

$$\delta_{ij} \int_0^h ds \int_0^s dt_1 \int_0^h dt_2\, \delta(t_1 - t_2) + \delta_{ij} \int_0^h ds \int_0^s dt_1 \int_s^h dt_2\, \delta(t_1 - t_2) = \frac{h^2}{2}\delta_{ij}$$

(31 h)

Here we have used the definitions (27 a) and (27 b) as well as (15 b) in the first two lines. In the last line, we have concluded that in the last term $t_1 < s < t_2$ is valid and therefore the delta functional vanishes, because $t_1 \neq t_2$. Thus only the first term in the last line remains.

Substituting (31 a) to (31 h) into (30) gives for $M_1$ and $M_2$ of (29):

$$M_1 = \partial_i M\, f^i + \tfrac{1}{2}\partial_i\partial_j M\, \sigma^{ik}\sigma^{jk},$$ (32 a)

$$M_2 = \tfrac{1}{2}\partial_i M \cdot \left(\partial_j f^i f^j + \partial_j\partial_k f^i \sigma^{jm}\sigma^{km}\right) +$$
$$\tfrac{1}{2}\partial_i\partial_j M \cdot \left(f^i f^j + 2\partial_k f^i \sigma^{kl}\sigma^{jl}\right) +$$
$$\tfrac{1}{2}\partial_i\partial_j\partial_k M \cdot \sigma^{il}\sigma^{jl} f^k +$$
$$\tfrac{1}{24}\partial_i\partial_j\partial_k\partial_l M \cdot \sigma^{im}\sigma^{jn}\sigma^{kr}\sigma^{ls}\left(\delta_{mn}\delta_{rs} + \delta_{mr}\delta_{ns} + \delta_{ms}\delta_{nr}\right).$$ (32 b)

This result is mainly a consequence of (31 a) to (31 h). We would have obtained the same averages (31), if we had started from the following simplified functionals:

$$F_i(h) := \tfrac{1}{2}h W_j(h),$$ (33 a)

$$G_{ij}(h) := \tfrac{1}{2}h W_i(h)W_j(h).$$ (33 b)

With (33), we sould have arrived at (32), too. Substituting (33) into (26) gives:

$$x^i(h) \approx x^i(0) + h f^i + \tfrac{1}{2}h^2 f^k \partial_k f^i + \sigma^{ij}W_j(h) + \frac{h}{2}W_l(h)\sigma^{kl}\partial_k f^i$$
$$+ \frac{h}{4}W_q(h)W_n(h)\partial_k\partial_m f^i \sigma^{kq}\sigma^{mn}$$ (34)

Hence this result reproduces an observable $\langle M(\vec{x}(h))\rangle$ to the same order in $h$ as (26) does, i.e. to an order including $h^2$. If we were only interested in the reproduction of that observable to an order including $h$, we would immediately get from (34):

$$\xi^i(h) := x^i(h) \approx x^i(0) + h f^i + \sigma^{ij}W_j(h),$$ (35)

i.e. an Euler algorithm, where no derivatives of $f^i$ occur. To get rid of the derivatives in (34), we have to reformulate it:

$$x^i(h) = x^i(0) + \sigma^{ij} W_j(h) +$$
$$\frac{h}{2}\left\{ f^i + \left[ f^i + \underbrace{\left(h f^k + \sigma^{kl} W_l(h)\right)}_{\xi^k(h) - x^k(h)} \partial_k f^i + \tfrac{1}{2} \partial_k \partial_m f^i \sigma^{kq} W_q(h) \sigma^{mn} W_n(h) \right] \right\}$$
$$\approx x^i(0) + \frac{h}{2}\left[ f^i(\vec{x}(0)) + f^i(\vec{\xi}(0)) \right] + \sigma^{ij} W_j(h) \tag{36}$$

In the second line, we have applied the Ito-formula (23) or (24) (together with (25)) to arrive at the last line (or vice versa), and have used (35) as a predictor step. (36) together with (35) is called a Heun algorithm.

$W_i(h)$ can be gaussian distributed random numbers that fulfil (31 b) and (31 c). In the next chapter, we are going to show that it is possible to construct $W_i(h)$ with uniformly distributed random numbers in an efficient way, avoiding gaussian distributed random numbers.

**Construction of the random variables** $W_i(h)$

Firstly, we address to the Euler algorithm (35) making an ansatz for $W_i(h)$:

$$W_i(h) := \begin{cases} a, & r_i < R \leq 1 \\ b, & r_i \geq R \end{cases} \tag{37}$$

with a uniformly distributed random number $r_i$ on the interval $[0,1]$. Since that algorithm is only valid to orders including $h$, just (31 a), (31 c) and (31 d) need to be fulfilled by (37). The average $\langle W_i(h) \rangle$ can be expressed by means of a probability density $\rho(r_i)$

$$\langle W_i(h) \rangle = \int_0^1 dr_i\, \rho(r_i) W_i(h) \tag{38}$$

with

$$\int_0^1 dr_i\, \rho(r_i) = 1, \tag{39}$$

but of course, for uniformly distributed numbers $r_i$, $\rho(r_i)$ ist just 1:

$$\rho(r_i) = 1. \tag{40}$$

Arbitrary powers of (37) read:

$$W_i^n(h) := \begin{cases} a^n, & r_i < R \leq 1 \\ b^n, & r_i \geq R \end{cases}$$

(41)

The average value of (41) can be expressed with the help of (38):

$$\langle W_i^n(h) \rangle = \int_0^R dr_i \, W_i^n(h) + \int_R^1 dr_i \, W_i^n(h) = a^n R + b^n (1-R)$$

. (42)

Together with (31 a), (31 c) and (31 d), we obtain a system of three equations for the three variables $a$, $b$ and $R$:

$$0 = \langle W_i(h) \rangle = a R + b (1-R),$$

(43 a)

$$h = \langle W_i^2(h) \rangle = a^2 R + b^2 (1-R),$$

(43 b)

$$0 = \langle W_i^3(h) \rangle = a^3 R + b^3 (1-R).$$

(43 c)

From (43 a), you can e.g. extract $b$ in terms of $a$ and $R$:

$$b = -a \frac{R}{1-R}.$$

(44 a)

Taking this result, you get $a$ in terms of $R$ from (43 b):

$$a = \pm \sqrt{\frac{1-R}{R} h} \quad \Rightarrow \quad b = \mp \sqrt{\frac{R}{1-R} h}.$$

(44 b)

With (44 b), we get from (43 c):

$$R = \tfrac{1}{2}.$$

(44 c)

Thus, the solution of (43) reads:

$$W_i(h) = \begin{cases} \pm \sqrt{h}, & r_i < \tfrac{1}{2} \\ \mp \sqrt{h}, & r_i \geq \tfrac{1}{2} \end{cases}$$

(45)

On the other hand, (45) can be reformulated in the subsequent way:

$$W_i(h) = A \sqrt{h} \left( r_j - \tfrac{1}{2} \right),$$

(46)

where $A$ is a normalization factor that can be calculated by means of (31 a), (31 c) and (31 d):

$$0 = \langle W_i(h) \rangle = \int_0^1 dr_i\, W_i(h) = A\sqrt{h} \left[ \tfrac{1}{2} r_i - \tfrac{1}{2} r_i \right]_0^h = 0 \quad , \tag{47 a}$$

$$0 = \langle W_i^3(h) \rangle \quad , \tag{47 b}$$

$$h = \langle W_i^2(h) \rangle = \int_0^1 dr_i\, W_i^2(h) = A^2 h \int_0^1 dr_i\, \left(r_i - \tfrac{1}{2}\right)^2 = \tfrac{1}{3} A^2 h \left[ y^3 \right]_{-\tfrac{1}{2}}^{\tfrac{1}{2}} = \frac{h}{12} A^2 \quad . \tag{47 c}$$

Thus $A = \pm\sqrt{12}$ and therefore we get the result

$$W_i(h) = \pm\sqrt{12h}\, \left(r_j - \tfrac{1}{2}\right) . \tag{48}$$

Hence for the Euler algorithm either (45) or (48) can be used. Unfortunately, they are no more valid for the Heun algorithm, because (31 e) cannot be fulfilled by them:

$\langle W_i^4(h) \rangle$ should be $3h^2$, but e.g. with (45) the former quantity becomes just $h^2$. Therefore, we have to choose a more complicated ansatz for $W_i(h)$ than (37):

$$W_i(h) = \begin{cases} a, & r_i \leq R_1 \\ b, & R_1 < r_i \leq R_2 \\ c, & r_i > R_2 \end{cases} . \tag{49}$$

And this time, (31 a), (31 c), (31 d), (31 e) and (31 f) must be fulfilled by (49). An arbitrary power of $W_i(h)$ reads:

$$W_i^n(h) = \begin{cases} a^n, & r_i \leq R_1 \\ b^n, & R_1 < r_i \leq R_2 \\ c^n, & r_i > R_2 \end{cases} . \tag{50}$$

Its average therefore becomes

$$\langle W_i^n(h) \rangle = \int_0^1 dr_i\, W_i^n(h) = a^n R_1 + b^n (R_2 - R_1) + c^n (1 - R_2) \quad . \tag{51}$$

The five parameters $a$, $b$, $c$, $R_1$ and $R_2$ can be extracted from the following system of five equations

$$0 = \langle W_i(h) \rangle = a R_1 + b(R_2 - R_1) + c(1 - R_2) \quad , \tag{52 a}$$

$$h = \langle W_i^2(h) \rangle = a^2 R_1 + b^2 (R_2 - R_1) + c^2 (1 - R_2) \quad , \tag{52 b}$$

$$0 = \langle W_i^3(h) \rangle = a^3 R_1 + b^3 (R_2 - R_1) + c^3 (1 - R_2) \quad , \tag{52 c}$$

$$3h^2 = \langle W_i^4(h)\rangle = a^4 R_1 + b^4 (R_2 - R_1) + c^4 (1 - R_2), \quad (52\text{ d})$$

$$0 = \langle W_i^5(h)\rangle = a^5 R_1 + b^5 (R_2 - R_1) + c^5 (1 - R_2), \quad (52\text{ e})$$

where (51) was used in combination with (31 a,c,d,e,f). To simplify these equations, we set

$$b := 0. \quad (53)$$

Then we get from (52 a):

$$c = -a \frac{R_1}{1 - R_2}. \quad (54)$$

Using (53) and (54) in (52 b), we obtain:

$$a = \pm \sqrt{h} \sqrt{\frac{1 - R_2}{R_1(1 + R_1 - R_2)}} \quad \Rightarrow \quad c = \mp \sqrt{h} \sqrt{\frac{R_1}{(1 - R_2)(1 + R_1 - R_2)}}. \quad (55)$$

Substituting (53) and (55) into (52 c) and (52 e), gives:

$$\left[\frac{1 - R_2}{R_1(1 + R_1 - R_2)}\right]^3 R_1^2 = \left[\frac{R_1}{(1 - R_2)(1 + R_1 - R_2)}\right]^3 (1 - R_2)^2, \quad (56\text{ a})$$

$$\left[\frac{1 - R_2}{R_1(1 + R_1 - R_2)}\right]^5 R_1^2 = \left[\frac{R_1}{(1 - R_2)(1 + R_1 - R_2)}\right]^5 (1 - R_2)^2. \quad (56\text{ b})$$

The division of (56 b) by (56 a) results into a quadratic equation for $R_1$ and $R_2$ with the solution

$$R_1 = 1 - R_2. \quad (57)$$

Finally, (53), (55) and (57) are used in (52 d) to obtain

$$R_1 = \frac{1}{6} \quad \Rightarrow \quad R_2 = \frac{5}{6}. \quad (58)$$

Thus, (49) now reads:

$$W_i(h) = \begin{cases} \pm \sqrt{3h}, & r_i \leq 1/6 \\ 0, & 1/6 < r_i \leq 5/6 \\ \mp \sqrt{3h}, & r_i > 5/6 \end{cases}. \quad (59)$$

Hence (59) can be used for the Heun-algorithm.

Now that we have numerical algorithms for solving the stochastic equations (12) taking into account (15), we want to return to physics, especially to (1), in order to show that the Heun-algorithm leads to a velocity-Verlet-like algorithm for it.

**Velocity-Verlet-like algorithm for stochastic dynamics**
The Langevin-type of equations for a stochastic motion (1) was expressed by (12) with the aid of (13). A numerical solution of (12) can be found by means of the Heun-algorithm (36) together with (35). For the predictor and corrector steps (35) and (36), respectively, we therefore get:

$$(\xi_i(h)) = \begin{pmatrix} \vec{\xi}_i(h) \\ m_i \vec{\omega}_i \end{pmatrix} = \begin{pmatrix} \vec{r}_i(0) + \vec{v}_i(0)h \\ m_i \vec{v}_i(0) + [\vec{F}_i(0) - m_i \gamma \vec{v}_i(0)]h + \sigma_i \vec{W}_i(h) \end{pmatrix}, \quad (60)$$

$$(x_i(h)) = \begin{pmatrix} \vec{r}_i(h) \\ m_i \vec{v}_i(h) \end{pmatrix} = \begin{pmatrix} \vec{r}_i(0) + \tfrac{1}{2} h[\vec{v}_i(0) + \vec{\omega}_i(h)] \\ m_i \vec{v}_i(0) + \tfrac{1}{2} h[(\vec{F}_i(0) - m_i \gamma \vec{v}_i(0)) + (\vec{F}_i(\vec{\xi}(h)) - m_i \gamma \vec{\omega}_i(h))] + \sigma_i \vec{W}_i(h) \end{pmatrix}. \quad (61)$$

Substituting $\vec{\omega}_i(h)$ from (60) into $\vec{r}_i(h)$ from (61), we obtain:

$$\vec{r}_i(h) = \vec{r}_i(0) + \vec{v}_i(0)h\left(1 - \frac{\gamma}{2}h\right) + \tfrac{1}{2} h^2 \frac{\vec{F}_i(0)}{m_i} + \tfrac{1}{2} \sigma_i \frac{h}{m_i} \vec{W}_i(h). \quad (62)$$

The term $\tfrac{1}{2} h[\vec{v}_i(0) + \vec{\omega}_i(h)]$ in the last (three) components of (61) can be replaced by the first (three) components of (61):

$$m_i \vec{v}_i(h) = m_i \vec{v}_i(0) + \tfrac{1}{2} h[\vec{F}_i(0) - \vec{F}_i(\vec{\xi}(h))] - m_i \gamma [\vec{v}_i(0) + \vec{\omega}_i(h)]\tfrac{1}{2} h + \sigma_i \vec{W}_i(h)$$
$$= m_i \vec{v}_i(0) + \tfrac{1}{2} h[\vec{F}_i(0) - \vec{F}_i(\vec{r}(h))] - m_i \gamma [\vec{r}_i(h) + \vec{r}_i(0)] + \sigma_i \vec{W}_i(h) \quad (63)$$

In (63), $\vec{F}_i(\vec{\xi}(h))$ could be replaced by $\vec{F}_i(\vec{r}(h))$, too, because $\vec{\xi}_i(h)$ stems from the calculation of the particle positions based on the Euler-algorithm and is therefore correct to an order including $h$, whereas $\vec{r}_i(h)$ does the same, but is correct to an order including $h^2$. Using a more exact determination algorithm for the positions should not deteriorate the exactness of the Heun-algorithm.

The three components of $W_i(h)$ can be calculated with the help of (59):

$$[\vec{W}_{i=1,\ldots,N}(h)]_{k=1,2,3} = \vec{W}_{i=1,\ldots,N}(h). \quad (64)$$

The ordinary velocity-Verlet algorithm for solving the newtonian equations of motion (7) follows from (62) and (63) for a vanishing damping constant $\gamma$ (and a therefore vanishing $\sigma_i$):

$$\vec{r}_i(h) = \vec{r}_i(0) + \vec{v}_i(0)h + \tfrac{1}{2} h^2 \frac{\vec{F}_i(0)}{m_i}, \quad (65\text{ a})$$

$$m_i \vec{v}_i(h) = m_i \vec{v}_i(0) + \tfrac{1}{2} h[\vec{F}_i(0) - \vec{F}_i(\vec{r}(h))]. \quad (65\text{ b})$$

**Scheme to maintain the conservation of the total linear momentum**

Als already described in the introduction of this text, the presence of the stochastic forces in the equations of motion of the particles destroys the property of the total linear momentum of being conserved (whereas this property is valid for pure newtonian dynamics). But this feature is easily maintained by imposing the subsequent additional requirement on $\vec{W}_i(h)$:

$$\sum_{i=1}^{N} \vec{W}_i(h) = 0 \tag{66}$$

in connection with

$$\sum_{i=1}^{N} m_i \vec{v}_i(0) = 0, \tag{67}$$

i.e. the initial velocities $\vec{v}_i(0)$ are chosen in a way that the initial total linear momentum vanishes, too. This can be seen by summation over the particle index *i* of (63) and thus also of (62). Here we must take into account Newton's third law:

$$\sum_{i=1}^{N} \vec{F}_i(0) = 0 \tag{68 a}$$

$$\sum_{i=1}^{N} \vec{F}_i(\vec{r}(h)) = 0 \tag{68 b}$$

With (66) and (67) as well as (68),

$$\sum_{i=1}^{N} m_i \vec{v}_i(h) = 0 \tag{69}$$

is true for all time steps *h*. (69) thus guarantees that the particles' centre of mass is kept fixed during the whole simulation with the velocity-Verlet-like algorithm (62), (63).

(66) means a further requirement being imposed on the random numbers $W_i(h)$ in (49) which has an impact on their choice, because we must be looking for new $\widetilde{W}_i(h)$ not only fulfilling (52 a,b,c,d,e) but also

$$\sum_{i=1}^{N} \widetilde{W}_i(h) = 0. \tag{70}$$

The connection between $\widetilde{W}_i(h)$ and $W_i(h)$ is the following:

$$\sum_{i=1}^{N} W_i(h) := c\, N \qquad (71)$$

(where c is a constant) and therefore $\widetilde{W}_i(h)$ must be

$$\widetilde{W}_i(h) = W_i(h) - c \qquad (72)$$

to guarantee (70). (52 a) can be fulfilled by $\widetilde{W}_i(h)$:

$$\langle \widetilde{W}_i(h) \rangle = \langle W_i(h) \rangle - \langle c \rangle = \langle W_i(h) \rangle - \frac{1}{N}\sum_{i=1}^{N} \langle W_i(h) \rangle = 0, \qquad (73)$$

where we have replaced *c* by means of its definition (71). (52 c) and (52 e) are statisfied by (72), too, because of (31 d) and (31 f) which we demonstrate for the case of (52 c):

$$\langle \widetilde{W}_i^3(h) \rangle = \langle (W_i(h) - c)^3 \rangle = \langle W_i^3(h) \rangle - 3\langle W_i^2(h)c \rangle + 3\langle W_i(h)c^2 \rangle - \langle c^3 \rangle. \qquad (74\ a)$$

All the terms in (74 a) contain a product of three $W_i(h)$ and thus must be zero:

$$\langle W_i^2(h)c \rangle = \frac{1}{N}\sum_{j=1}^{N} \langle W_i^2(h) W_j(h) \rangle = 0, \qquad (74\ b)$$

$$\langle W_i(h)c^2 \rangle = \frac{1}{N^2}\sum_{j,k=1}^{N} \langle W_i(h) W_j(h) W_k(h) \rangle = 0, \qquad (74\ c)$$

$$\langle c^3 \rangle = \frac{1}{N^3}\sum_{j,k,l=1}^{N} \langle W_j(h) W_k(h) W_l(h) \rangle = 0. \qquad (74\ d)$$

The analogue is true for (52 e), because then all terms contain products of five $W_i(h)$.
Unfortunately, neither (52 b) nor (52 d) can be satisfied by $\widetilde{W}_i(h)$:

$$\langle \widetilde{W}_i^2(h) \rangle = \langle (W_i(h) - c)^2 \rangle = \underbrace{\langle W_i^2(h) \rangle}_{h} - 2\langle W_i(h)c \rangle + \langle c^2 \rangle = h\left(1 - \frac{1}{N}\right) \qquad (75)$$

due to

$$\langle W_i(h)c \rangle = \frac{1}{N}\sum_{j=1}^{N} \langle W_i(h) W_j(h) \rangle = \frac{h}{N}\sum_{j=1}^{N} \delta_{ij} = \frac{h}{N},$$

$$\langle c^2 \rangle = \frac{1}{N^2}\sum_{j,k=1}^{N} \langle W_j(h) W_k(h) \rangle = \frac{h}{N^2}\sum_{j,k=1}^{N} \delta_{jk} = \frac{h}{N}$$

and

$$\langle \widetilde{W}_i^4(h) \rangle = \langle (W_i(h)-c)^4 \rangle = \underbrace{\langle W_i^4(h) \rangle}_{3h^2} - 4\langle W_i^3(h)c \rangle + 6\langle W_i^2(h)c^2 \rangle - 4\langle W_i(h)c^3 \rangle + \langle c^4 \rangle$$

$$= 3h^2\left(1-\frac{1}{N}\right)^2 \tag{76}$$

because of

$$\langle W_i^3(h)c \rangle = \frac{3}{N}h^2,$$
$$\langle W_i^2(h)c^2 \rangle = \frac{2+N}{N^2}h^2,$$
$$\langle W_i(h)c^3 \rangle = \frac{3}{N^2}h^2,$$
$$\langle c^4 \rangle = \frac{3}{N^2}h^2.$$

Both (75) and (76) differ from (52 b) and (52 d), respectively, by powers of the same factor $\left(1-\frac{1}{N}\right)$ which might be neglectable for large particle numbers $N$. But even for small numbers of particles, this factor poses no problems, because it can be absorbed into the definition of $\sigma_j$, see equation (4):

$$\sigma_j \to \sigma_j \cdot \sqrt{1-\frac{1}{N}}. \tag{77}$$

With this prescription for the choice of $\sigma_j$, the $W_i(h)$ from (59) may be used in connection with (66).